\documentclass[aps,prl,twocolumn,psfig,showpacs]{revtex4}
\usepackage{epsfig}

\input {epsf.sty}
\begin{document}
\draft
\title{Composite-fermion description of rotating Bose gases at low angular momenta}
\author{M.N. Korslund and S. Viefers}
\address{Department of Physics,  University of Oslo,
P.O. Box 1048 Blindern, N-0316 Oslo, Norway }
\date{\today}
\begin{abstract}
We study the composite fermion construction at and below the single vortex ($L=N$) 
state of weakly interacting rotating Bose gases, presenting a new method
for handling the large number of derivatives typically occurring via the Slater 
determinant.
Remarkably, the CF wave function at $L=N$ becomes asymptotically {\em exact} in the 
thermodynamic
limit, even though this construction is not, {\em a priori}, expected to work
in the low angular momentum regime. This implies an interesting mathematical
identity which may be useful in other contexts.

\end{abstract}
\maketitle
\vspace{-11pt}

\newcommand{\half}{\frac 1 2 }
\newcommand{\eg}{{\em e.g.} }
\newcommand{\ie}{{\em i.e.} }
\newcommand{\etc} {{\em etc.}}

\newcommand{\noi}{\noindent}
\newcommand{\etal}{{\em et al.\ }}
\newcommand{\cf}{{\em cf. }}

\newcommand{\dd}[2]{{\rmd{#1}\over\rmd{#2}}}
\newcommand{\pdd}[2]{{\partial{#1}\over\partial{#2}}}
\newcommand{\pa}[1]{\partial_{#1}}
\newcommand{\pref}[1]{(\ref{#1})}

\newcommand{\be}{\begin{eqnarray}} 
\newcommand{\ee}{\end{eqnarray}}
\newcommand{\e}{\varepsilon} 
\newcommand{\D}{\partial}
\newcommand{\pt}{\tilde p}
\newcommand{\p}{\partial}
\newcommand{\jas}{\prod_{i<j}(z_i - z_j)}




\section{Introduction}
Rotating atomic Bose condensates have been subject to ever increasing theoretical and
experimental interest in recent years\cite{baym1}. On the experimental side, 
substantial progress has been made in increasing the rotation such as to produce 
ever larger amounts of vortices \cite{madison1, abo1, haljan1,engels1}
and, very recently, coming close to the point where the vortex lattice is
expected to melt and the system is expected to enter the quantum Hall 
regime\cite{schweickhard1}.
Theoretical studies have addressed the entire range of angular momenta, from
few-vortex states ($L \sim N$ where $L$ is the total angular momentum and $N$ the
number of particles), up to the regime $L \sim N^2$ where the system 
displays a number of close analogies to the fractional quantum Hall
 effect (FQHE)\cite{wilkin1, wilkin2, cooper1}.

A theoretical tool which has been very successful in describing the physics in
the quantum Hall regime is the composite fermion (CF) construction\cite{jainreview}. Originally
developed in the context of the fractional quantum Hall effect, CF wave functions
have, more recently, been applied to the high angular momentum states
of weakly interacting Bose gases\cite{cooper1, viefers1, chang1, regnault1}. 
By construction, this method has been 
assumed to be applicable only in the FQH regime, $L\sim N^2$. However, as
was pointed out in \cite{viefers1}, there were some indications that the CF wave functions
might do surprisingly well even at the lowest angular momenta. In this work we thus
discuss the CF construction in the regime $0 \leq L \leq N$. In particular, we present
a numerical study which shows that the CF wave function for the single vortex
($L=N$) becomes {\em exact} in
the large $N$ limit. One of our main motivations for studying this case in 
more detail were earlier calculations by Cooper {et al} \cite{cooper1}, which revealed 
an increase in the overlap as function of $N$ for up to 10 particles.
We confirm that this trend continues in a systematic way for higher $N$, 
with the deviation of the overlap from unity decreasing as $\sim 1/N$. 
Since the exact ground state wave function is known analytically in this
case \cite{wilkin1, bertsch99,smith1}, this result suggests a non-trivial mathematical identity,
valid asymptotically in the limit $N \rightarrow \infty$, which
may turn out useful in other contexts. Our solution involves a new way
of handling the large number of derivatives typically occurring in the CF construction
when going to low angular momenta. This large number of derivatives  has been
one of the main obstacles to taking the CF construction to higher particle numbers
than, say, 10 \cite{jaincomm}. We have computed the overlap
of the single vortex CF wave function with the exact one for up to 43 
particles, which is a far higher number than could be achieved previously.
This was done by taking advantage of the mathematical properties of the determinants
involved, which enabled us to rewrite the wave functions in a simpler way,
without derivatives, before performing the numerical calculation of the overlaps.
A straightforward modification of this method allowed us to study other angular
momenta as well, and revealed some peculiar mathematical properties of the
CF construction. We would like to emphasize that this is mainly a technical
work in the sense that our main emphasis is on this new method and
on the mathematical properties of the CF construction. While our study was 
performed in an angular momentum regime where exact analytical wave 
functions are known, we hope that our results and methods may turn 
out useful to study states beyond the single vortex, where this is not the
case. For example, the same approach may be used to study the case $L=2N$,
where the system is expected to be close to the transition between a two- and
a three-vortex state. 
Our preliminary findings indicate that the aforementioned
mathematical structures may greatly simplify the CF study of this and other
few-vortex states.

The paper is organized as follows: We start by reviewing the construction of CF 
trial wave functions in the context of rotating hard-core bosons, illustrated by some
simple examples. We then present evidence
that the overlap of the single vortex CF wave function with the exact one converges
towards 1 in the large $N$ limit. Since we believe that (modified versions of) our approach
may provide a useful tool for studying other CF wavefunctions with a high number
of derivatives, our method of handling the derivatives is explained in some detail.
In the following section, we discuss the region $L < N$. The CF wave functions are,
in general, not exact here but do reproduce the exact wave function for $L=0,2,3$
which, again, suggests some interesting mathematical identities. Finally, we
summarize our work.

\section{CF description of rotating Bose systems}
Consider a system of $N$ spinless bosons with mass $m$
in a harmonic trap of strength $\omega$,
rotating with angular frequency $\Omega$ 
and interacting via a short-range (delta function) potential $H_I$. In a rotating frame
the Hamiltonian can be written as
\be
H = \sum_{i=1}^N \left[\frac{\vec p_i^2}{2m} + \frac 1 2 m \omega^2 \vec r_i^2 \right]
-\Omega L_z + H_I
\ee 
where  $L_z$ denotes
the angular momentum around the rotation axis.
Reshuffling a few terms, this is straightforwardly rewritten as
\be
H = \sum_{i=1}^N \left[ \frac{1}{2m}\left( \vec p_i - \vec A \right)_{\parallel}^2  
+ H_{ho}(z_i)\right] 
+(\omega -\Omega) L_z + H_I
\ee
with $\vec A = m\omega (-y,x)$, $H_{ho}(z)$ denoting the $z$-part of the harmonic 
oscillator Hamiltonian and $\parallel$
denoting the planar ($x,y$) part of the Hamiltonian. This is how the formal link
to the quantum Hall system comes about: We see that the planar part of $H$
takes the form of particles moving in an effective  "magnetic" field
$\vec B_{eff}=  \nabla \times \vec A = 2m\omega \hat z$.
Now, the interaction is assumed to be weak in the sense that it does not mix different
harmonic oscillator levels. We will be interested, for a given total angular momentum, 
only in the {\em lowest} many-body states (the "yrast" band).
In this limit, the model may be rewritten as a lowest
Landau level (LLL) problem in the effective "magnetic" field
$B_{eff}= 2m\omega$ (and of course, $n_z=0$ for the harmonic oscillator in the
 $z$-direction). The Hamiltonian then takes the form
\be
H = (\omega - \Omega)  L + g\sum_{i<j} \delta^2(\bf r_i - \bf r_j)
\ee
($\hbar=1$)
where we now use $L$ to denote the total angular momentum,
$L=\sum_i l_i = L_z$.

The single particle states spanning our Hilbert space (the lowest Landau level)
are 
\be
\eta_{0,l} = \frac{1}{\sqrt{2^{l+1}\pi l!}} z^l e^{-\bar z z/4}
\ee
where $z = \sqrt{2m\omega} (x+iy)$ are complex coordinates denoting the
particle positions in the plane, and $l$ is the angular momentum of the state.
A general bosonic many-body wave function $\psi(z_1,...z_N)$ 
will then be a homogeneous, symmetric polynomial in the $z_i$:s, times the
exponential factor $exp(-\sum_i |z_i|^2/4)$ 
(which will be suppressed throughout this paper for simplicity). The degree of the
polynomial gives the total angular momentum of the state. A special class of
such wave functions are the so-called {\em composite fermion} (CF) wave functions.
They were first introduced by Jain\cite{jainreview} and have been extremely successful
in describing FQH states, quantum dots in high magnetic fields \cite{qdcf} and, more recently,
highly rotational states of Bose condensates\cite{cooper1, viefers1, chang1, regnault1}. 
In quantum Hall physics, the 
main idea of this construction is, roughly speaking, to attach an even number of flux
quanta to each electron, thus mapping them into weakly interacting {\em composite}
fermions which can be thought of as moving in a reduced magnetic field. 
Technically, "attaching a flux quantum" means multiplying the wave
function by a Jastrow factor,
\be
\prod_{i<j} (z_i - z_j).
\ee
We see that the Jastrow factor has the effect of keeping the particles apart
-- it goes to zero if any two coordinates $z_i$ and $z_j$ approach each other.
Therefore, it "takes care of" much of the repulsive interaction between the
particles.
In the simplest approach, the so-called non-interacting composite fermion 
(NICF) approach, the wave function is then simply constructed as a Slater
determinant of (non-interacting) composite fermions in the reduced magnetic
field, times an even power of Jastrow factors. In the case of bosons, whose wave function
has to be symmetric rather than antisymmetric, one instead absorbs an {\em odd}
power of flux quanta, mapping the bosons to weakly interacting composite
fermions. In other words, bosonic trial wave functions 
with angular momentum $L$ are constructed as non-interacting fermionic
wave functions with angular momentum $L - mN(N-1)/2$, multiplied by an {\em odd}
number  $m$ of Jastrow factors ($m=1$ throughout this paper), 
and projected onto the LLL,
\be
\psi_L = {\cal P}\, \left(  f_S(z_i, \bar z_i) \jas^m \right).
\label{cfwf}
\ee
Here, $f_S$ denotes a Slater determinant consisting of single-particle wave functions
$\eta_{nl}(z, \bar z) \propto z^l L_n^l(z \bar z/2)$ where $n$ is the (CF) Landau
level index ($l \geq -n$) and $L_n^l$ a generalized Laguerre polynomial. The LLL
projection $\cal P$ amounts to the replacement $\bar z_i \rightarrow 2\partial/\partial z_i$ in
the polynomial part of the wave function --
the recipe is to replace all $\bar z$:s
with derivatives in the final polynomial, after multiplying out the Slater determinant
and the Jastrow factors and moving all $\bar z$:s to the left. 
It has been shown\cite{jainreview} that with this projection method, 
the single-particle wave 
functions in the CF Slater determinant may be written as
\be
\eta_{nl} = z^{n+l} \partial^n, ~~~~ l \geq -n
\label{etanl}
\ee
with all derivatives acting only to the right.
As this method tends
to get computationally heavy in numerical calculations with many particles and
a large number of derivatives, slightly different methods of obtaining 
LLL wavefunctions
have been employed in most of the CF literature\cite{jainreview}. These, too,
are often referred to as projection.
Nevertheless, in this paper,
"projection" will refer to the above brute force procedure.
 
 Before moving on to low angular momenta, let us illustrate the method on two
 simple and well-known examples in the QH regime: 
 First, consider the case $L = N(N-1)$. Taking $m=1$, the
 Slater determinant $f_S$ has to contribute an angular momentum
 $N(N-1)/2$ and is
 given by putting all CFs into the lowest CF Landau level, from $l=0$ to
 $l= N-1$,
\be
f_S=
 \left| \begin{array}{cccc}
1 & 1 &  ... & 1 \\
z_1 & z_2 & ... & z_N \\
z_1^2 & z_2^2 & ... & z_N^2 \\
... & ... & ... & ... \\
z_1^{N-1}& z_2^{N-1} & ... & z_N^{N-1}
 \end{array} \right|
 \equiv \jas.
 \ee
 We immediately see from Eq.\pref{cfwf} that the full wave function is simply the bosonic
 Laughlin wave function 
 \be
 \psi_{L} = \jas^2
 \label{psil}
 \ee
 with angular momentum $L = N(N-1)$. This is the {\em exact} ground state for
 the delta function interaction.
 Next, consider the angular momentum 
 $N(N-1) - N$, corresponding to a bosonic quantum Hall "quasiparticle" (as opposed to
 quasihole). In order to {\em decrease} the angular momentum by $N$ 
 as compared to the Laughlin state, we need to move one CF to the second 
 Landau level, {\em i.e.} construct the Slater determinant
 \be
f_S=
 \left| \begin{array}{cccc}
\bar z_1 & \bar z_2 & ... & \bar z_N \\
1 & 1 &  ... & 1 \\
z_1 & z_2 & ... & z_N \\
z_1^2 & z_2^2 & ... & z_N^2 \\
... & ... & ... & ... \\
z_1^{N-2}& z_2^{N-2} & ... & z_N^{N-2}
 \end{array} \right|.
 \ee
 This gives the full trial wave function (again, apart from the exponential factor)
 \be
 \psi_{qp} &=& \sum_{i=1}^N (-1)^i  \partial_i \prod_{k<l ; k,l \neq i} (z_k - z_l) \prod_{m<n}^N (z_m - z_n)
 \nonumber \\
 	        &\propto& \sum_{i=1}^N \sum_{j\neq i} \frac{1}{z_i - z_j} \prod_{k\neq i} (z_i - z_k)^{-1} \, \psi_L 
 \ee
 with $\psi_L$ denoting the Laughlin state \pref{psil}.
 This wave function has very high overlap with the exact one
 ({\em e.g.}, 99.7 \% for 4 bosons \cite{viefers1}). Its fermionic counterpart has been proven to 
 capture correctly both the fractional charge and the anyonic statistics of the QH
 quasielectron\cite{kjonsberg1, jeon1}, and the same is expected to be the case
 for this bosonic version.
 
 Trial wave functions for other yrast states are constructed in similar ways. The lower
 the angular momentum, the larger the number of derivatives.

\section{Single vortex, $L=N$}
\begin{figure}[htp]
\centering
{\psfig{figure=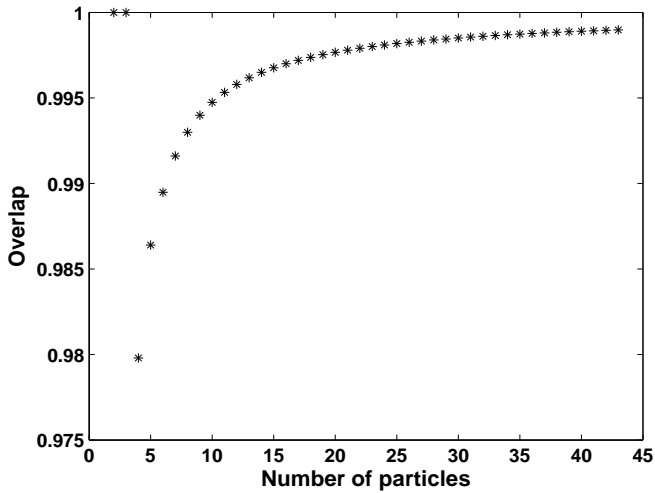, height=7.0cm, width=9.5cm,angle=0}}
\caption{The overlap between the $L=N$ CF trial wave function \pref{svcf} and the exact 
one \pref{svex} as function of the number of particles.}
\label{fig:overlap}
\end{figure}

In the examples above, we considered angular momenta $L \sim N^2$.
Let us now turn to the regime $L \sim N$ which, for large systems,
corresponds to much smaller angular momenta. Note that since the Jastrow
factor in Eq.\pref{cfwf} has itself an angular momentum of $mN(N-1)/2$, 
{\em we need to act with ${\cal O}(N^2)$ derivatives} in order to get down
to $L \sim N$. One would expect that these derivatives acting on
the Jastrow factor destroy most of the good correlations which are at the
very heart of the CF construction, which would thus fail
in this regime. Surprisingly, however, we shall see that this is not the case.
As shown in \cite{viefers1}, the Slater determinant for the ground state trial 
wave function at $L=N$ is constructed by occupying the single-particle
states $\eta_{n,-n}$ for $n=N-2$ through $n=0$, and the state $\eta_{0,1}$,
\be
f_S=
 \left| \begin{array}{cccc}
z_1 & z_2 & ... & z_N \\
1 & 1 &  ... & 1 \\
\bar z_1 & \bar z_2 & ... & \bar z_N \\
\bar z_1^2 & \bar z_2^2 & ... & \bar z_N^2 \\
... & ... & ... & ... \\
\bar z_1^{N-2}& \bar z_2^{N-2} & ... & \bar z_N^{N-2}
 \end{array} \right|
\label{svfs}
\ee
where, again, the $\bar z$:s are to be replaced by derivatives acting on the
Jastrow factor. Thus, the trial wave function may be written as \cite{viefers1}
\be
\psi_{(L=N)} = \sum_{n=1}^N (-1)^n z_n \prod_{k<l ; k,l \neq n} (\p_k -\p_l) \jas
\label{svcf}
\ee
On the other hand, it has been shown\cite{wilkin1, bertsch99, smith1} that the {\em exact}  wave
function for $L=N$, the single vortex, is given by
\be
\psi_{(L=N)}^{ex} = \prod_{i=1}^N (z_i - Z)
\label{svex}
\ee
where $Z = \sum_i z_i/N$ is the center of mass. It was shown in Ref. \cite{cooper1}
that, surprisingly, the overlap between \pref{svcf} and \pref{svex} appeared to increase with
increasing particle number (but only results up to $N=10$ were available at the time).
Here we have studied this point systematically and computed the overlaps for up to
43 particles. The mathematical approach which enabled us to handle such large
numbers of derivatives, will be described shortly. First, however, let us state the 
result: {\em The overlap between the CF trial wave function \pref{svcf} and the
exact one \pref{svex} converges towards 1 in the large $N$ limit}, see fig. \ref{fig:overlap}.
The overlap equals $98.64\%$ already at $N=5$ and $99.47\%$ for $N=10$. 
For $N=43$ it equals $99.90\%$. We have
checked that this is not a trivial result in the sense that the two wave functions 
might simply share the same leading term and all other terms become irrelevant 
for large $N$. Rather, we found that the leading term of the two wave functions, 
$\prod_i z_i$, only corresponds to a weight of about $50\%$
of the total wave function, and that the subleading terms of the two wave functions
converge towards each other as well. This suggests that
the functions \pref{svcf} and \pref{svex} {\em are identically equal to each other in
the limit} $N \rightarrow \infty$. This is a non-trivial mathematical identity which
we have so far not been able to prove analytically; however, we noticed by simple
curve-fitting, that the deviation of the overlap from unity decreases as $\sim1/N$. 
We will discuss other, similar identities in the section on $L < N$.

\subsection{Combinatorics and numerical method}
For the sake of clarity, we will illustrate our approach
to the single vortex wave function for the simple case $N=4$.
The generalization to arbitrary $N$ is straightforward, and
the general result will be stated at the end.
After projection, the CF Slater determinant \pref{svfs} takes
the form
\begin{equation}
\label{eqn:fS_determinant}
f_S = \begin{array}{|cccc|}
z_1 & z_2 & z_3 & z_4 \\
1 & 1 & 1 & 1 \\
\partial_1 & \partial_2 & \partial_3 & \partial_4 \\
\partial_1^2 & \partial_2^2 &\partial_3^2 & \partial_4^2
\end{array}\;.
\end{equation}
This determinant may be rewritten as a permutation sum,
\begin{equation}
f_S = \sum_\pi \epsilon(\pi)\, z_{\pi(1)}^1 z_{\pi(2)}^0
\partial_{\pi(3)}^1 \partial_{\pi(4)}^2
\label{fs}
\end{equation}
with $\pi$ denoting all permutations over the particle indices
(1,2,3,4), and $\epsilon(\pi)= \pm 1$ for even/odd permutations.
Similarly, we rewrite the Jastrow factor as
\begin{equation}
J = \begin{array}{|cccc|}
1 & 1 & 1 & 1 \\
z_1 & z_2 & z_3 & z_4 \\
z_1^2 & z_2^2 & z_3^2 & z_4^2 \\
z_1^3 & z_2^3 & z_3^3 & z_4^3
\end{array}
= \sum_\sigma \epsilon(\sigma)\, z_{\sigma(1)}^0 z_{\sigma(2)}^1
z_{\sigma(3)}^2 z_{\sigma(4)}^3.
\label{Jpsum}
\end{equation}
Now, the first crucial observation is that an equivalent way
of writing the permutation sum \pref{Jpsum} is by fixing the 
particle indices and summing over all permutations 
of the exponents instead,
\begin{equation}
J = \sum_\sigma \epsilon(\sigma)\, z_1^{\sigma(1)-1} z_2^{\sigma(2)-1}
z_3^{\sigma(3)-1} z_4^{\sigma(4)-1}.
\end{equation}
Moreover, since reordering the terms will leave the sum unchanged, 
we may apply any permutation $\pi$ to all the indices as long as 
we make sure to change the signs accordingly:
\begin{equation}
J = \sum_\sigma \epsilon(\pi\sigma)\, z_{\pi(1)}^{\sigma(1)-1} z_{\pi(2)}^{\sigma(2)-1}
z_{\pi(3)}^{\sigma(3)-1} z_{\pi(4)}^{\sigma(4)-1}.
\end{equation}
Noting that $\epsilon(\pi) \epsilon(\pi\sigma) = \epsilon(\sigma)$,
we can now combine the Slater determinant $f_S$ and the Jastrow 
factor $J$ to give the following expression for the (unnormalized)
single vortex wave function,
\be
\psi_{(L=N)} &=& \sum_\pi \sum_\sigma \epsilon(\sigma)\, z_{\pi(1)}^1
z_{\pi(2)}^0 \partial_{\pi(3)}^1 \partial_{\pi(4)}^2 \nonumber \\
&\times&\;\, z_{\pi(1)}^{\sigma(1)-1} z_{\pi(2)}^{\sigma(2)-1}
z_{\pi(3)}^{\sigma(3)-1} z_{\pi(4)}^{\sigma(4)-1}.
\ee
We note that the outer sum is simply a (symmetric) sum over all
permutations $\pi$ of the particles indices. Therefore, it is sufficient 
to study one term, denoted as $f_U$, in this permutation sum 
and perform the symmetrization only at the very end.
Choosing this term to be $\pi = (1,2,3,4)$, we get
\be
f_U &=& \sum_\sigma \epsilon(\sigma)\, z_1\; \partial_3\; \partial_4^2
\times z_1^{\sigma(1)-1} z_2^{\sigma(2)-1} z_3^{\sigma(3)-1}
z_4^{\sigma(4)-1} \nonumber \\
  &=& \sum_\sigma \epsilon(\sigma)\, z_1^{\sigma(1)}\;
z_2^{\sigma(2)-1}\; (\sigma(3)-1)z_3^{\sigma(3)-2}\; \nonumber \\
  &\times& (\sigma(4)-1)(\sigma(4)-2)z_4^{\sigma(4)-3} \; .
\label{eqn:sumMult}
\ee
The next important observation is that the above sum may
again be expressed as a determinant,
\begin{equation}
f_U = \begin{array}{|cccc|}
z_1   & 1     & 0     & 0\\
z_1^2 & z_2   & 1     & 0\\
z_1^3 & z_2^2 & 2z_3   & 2\\
z_1^4 & z_2^3 & 3z_3^2 & 6z_4\\
\end{array}\;.
\label{fu}
\end{equation}
Thus, to summarize, we have expressed both the Slater determinant
and the Jastrow factor in \pref{cfwf} as permutation sums, 
combined these, performed
all derivatives, and cast the final expression in the form of
a new determinant. The full wave function $\psi_{(L=N)}$ is 
obtained by adding all
permutation of the particle indices in \pref{fu}. We have obtained
an explicit form of the CF polynomial, without derivatives, which may
then be used in numerical calculations, to compute overlaps etc.

The generalization of \pref{fu} to arbitrary $N$ is straightforward;
one finds that
%
\begin{equation}
\label{eqn:matrixVortex}
f_U = \begin{array}{|ccccc|}
z_1   & 1     & 0      &
\cdots & 0\\
z_1^2 & z_2   & 1      &
\cdots & 0\\
\vdots & \vdots & \vdots & \cdots &
\vdots \\
z_1^{N-1} & z_2^{N-2} & {N-2 \choose 1}z_3^{N-3} & \cdots & 1 \\
z_1^N & z_2^{N-1} & {N-1 \choose 1}z_3^{N-2} &
\cdots& {N-1 \choose N-2}z_N
\end{array}\;.
\end{equation}
A particularly appealing property of this matrix is its 
near-triangular form, which greatly reduces the calculation 
cost.

Before ending this section, let us briefly outline how we 
used this result to compute overlaps between the CF wave function 
and the exact one \pref{svex}.
The polynomial part of $\psi_{(L=N)}$ may be regarded 
as a vector in the space of symmetric
polynomials. A convenient basis for this space consists of the 
symmetrized sums of
individual polynomial terms. For example, the symmetrization of
$z_1z_2z_3^2$ is

\begin{equation}
\label{eqn:symmetry}
{\cal S}[z_1z_2z_3^2] = 2z_1z_2z_3^2 + 2z_1z_2z_4^2 + 2z_1z_3z_2^2 + \dots
\end{equation}
and may be regarded as one basis vector. (The factor $2$ arises from
the fact that both eg. $z_1z_2z_3^2$ and $z_2z_1z_3^2$ occur in the
sum.) Since $\psi_{(L=N)} = {\cal S}[f_U]$ ($\cal S$ denoting symmetrization), 
it is therefore straightforward
to determine the representation of $\psi_{(L=N)}$ in the symmetric basis
by symmetrizing each term of $f_U$ individually.

Since overlap integrals between polynomial terms with non-matching powers
are zero, our symmetric basis is orthogonal. This
means that if both $\psi_{(L=N)}$ and $\psi_{(L=N)}^{\,ex}$ are
normalized and represented in a normalized symmetric basis, then the
overlap may be computed as a simple dot product. This also implies that
all calculated overlap values are exact (up to numerical precision).
To express $\psi_{(L=N)}^{\,ex}$ in this basis, let us rephrase the
product \pref{svex} as the sum
\begin{equation}
\psi_{(L=N)}^{\,ex} = \sum_{i=0}^N\frac{1}{(-N)^i} s_1^i\cdot s_{N-i},
\end{equation}
where $s_i$ = $s_i(z_1, \dots, z_N)$ are the fundamental symmetric
polynomials of degree $i$, 
\be
s_i(z_1, \dots, z_N) = {\cal S} \left[ z_1 z_2 ... z_i \right].
\ee
The products $s_N,\,$ $s_1s_{(N-1)}, \dots,
s_1^N$ appearing in this sum form another, non-orthogonal,
basis. Through a change of basis we can therefore convert
$\psi_{(L=N)}^{\,ex}$ to the orthogonal symmetrized basis,
enabling us to compute the overlaps.

The method described in this section may be straightforwardly
modified to other values of $L$. In the next section we illustrate how it may
be used to shed light on some yrast states at even lower angular momenta.

\section{Below the single vortex, $L < N $}
Exact ground state wave functions are known not only for the single vortex
\pref{svex} but for all angular momentum states $2 \leq L \leq N$. As was
shown some years ago\cite{bertsch99,smith1}, they are simply 
given by fundamental symmetric polynomials $s_L(\tilde z_i)$ where 
$\tilde z_i = z_i - Z$ and $Z$ is again the center-of-mass coordinate:
\be
\psi_L^{ex} = \sum_{p_1 < p_2 < ... < p_L} (z_{p_1} - Z)(z_{p_2} - Z) \cdots (z_{p_L} - Z).
\label{TI1}
\ee 
For example, $\psi_{L=2} = {\cal S}\left[(z_{1} - Z)(z_{2} - Z)\right] $, with $\cal S$ denoting
symmetrization over all particle coordinates. Note that these states are {\em translation invariant} (TI),
i.e. invariant under a simultaneous constant shift $z_i \rightarrow z_i + a$ of all the
coordinates\cite{viefers1,trugman1}. We see that the single vortex \pref{svex} is merely
a special case of this series of wave functions.
(At $L=0$ the polynomial part of the exact
ground state is trivially a constant as all particles are in the $l=0$ state, while
the only possible way of constructing the ground state at $L=1$ is $\psi_{L=1}= \sum_i z_i$,
which corresponds to a center of mass excitation of the $L=0$ state.)

As discussed in Refs. \cite{trugman1,viefers1}, the space of TI polynomials for $2 \leq L \leq N$ 
is spanned by the basis states
\be
| k_2 k_3 ... k_N\rangle =  s_2^{k_2}(\tilde z_i)\, s_3^{k_3}(\tilde z_i) \cdots s_N^{k_N}(\tilde z_i)
\ee
where $L = \sum_{n=2}^N nk_n$. We thus see that the states \pref{TI1} are special cases of
such basis states, with $k_L=1$, all other $k_n=0$. 
In anticipation of the discussion below, note that for $L=2,3$, they are the {\em only} 
possible basis states, and thus it is obvious that \pref{TI1} has to be exact for $L=2,3$.
(For larger $L$ this is no longer the case; for example, the basis for $L=4$ consists
of $|20000...\rangle$ as well as $|00100...\rangle$.)

Given these results, let us discuss some mathematical peculiarities 
of the CF construction in the regime $L < N$.
First of all, the CF construction immediately reproduces the exact wave
function for $L=0$: In order to cancel all the angular momentum of the
Jastrow factor, the CF Slater determinant with lowest possible (quasi)
Landau level energy is uniquely given by occupying the CF Landau level
states $\eta_{n,-n}$ for $n=0$ through $N-1$,
\be
f_S=
 \left| \begin{array}{cccc}
1 & 1 &  ... & 1 \\
\bar z_1 & \bar z_2 & ... & \bar z_N \\
\bar z_1^2 & \bar z_2^2 & ... & \bar z_N^2 \\
... & ... & ... & ... \\
\bar z_1^{N-1} & \bar z_2^{N-1} & ... & \bar z_N^{N-1} \\
 \end{array} \right|
 = \prod_{i<j} (\bar z_i - \bar z_j).
 \ee
 After projection, this will give a Jastrow factor of derivatives acting on the
 corresponding Jastrow factor of $z$:s, thus resulting in a constant. This can
 be made explicit using the notation of the previous section. Proceeding exactly as in 
 equations \pref{fs}-\pref{fu}, again taking $N=4$, we find that 
 $\psi_{L=0} = \sum_{\pi}f_U(\pi(1), \pi(2), \pi(3),\pi(4))$
with
\be
f_U&=& \sum_{\sigma} \e(\sigma) \p_1^0 \p_2^1 \p_3^2 \p_4^3 \times
                         z^{\sigma(1)}_1 z^{\sigma(2)}_2 z^{\sigma(3)}_3 z^{\sigma(4)}_4 \\
	&=&\sum_{\sigma} \e(\sigma) z_1^{\sigma(1)} \cdot \sigma(2)  z_2^{\sigma(2)-1} 
		\cdot \sigma(3)(\sigma(3) - 1)  z_3^{\sigma(3)-2}  \nonumber \\
		&\cdot& \sigma(4)(\sigma(4) - 1)(\sigma(4)-2)  z_4^{\sigma(4)-3} 
\ee
where $\sigma$ now denotes all permutations over (0,1,2,3). This is identified with the
determinant
\be
 \left| \begin{array}{cccc}
1 & 0 &  0 & 0 \\
 z_1 & 1 & 0 & 0 \\
z_1^2 & 2z_2 & 2 & 0 \\
z_1^3 & 3z_2^2 & 6z_3 & 6 \\
 \end{array} \right|
 = 12
 \ee
which obviously gives the correct wave function (up to normalization) 
for $L=0$ (all bosons in the $l=0$ state).

The CF construction reproduces the {\em exact} ground state wave functions for
$L = 2$ and $3$ as well. In contrast to the case $L=N$, these identities 
are not asymptotic -- they are exact for {\em all} $N$. 
This is not surprising, for the following reason\cite{viefers1}:
For these angular momenta, one can construct {\em compact}
CF wave functions which are thus known to be translation invariant 
(TI)\cite{footnote}.
On the other hand, as discussed above, for these lowest angular momenta 
there only exists one basis state in the space of TI polynomials. 
Therefore, the CF compact state has to correspond to this basis state,
which is the exact wave function.

What {\em is} somewhat surprising, however, is how this comes about mathematically.
Consider again the $L=0$ state discussed above. From this we can construct
compact $L=2$ states by moving {\em any} of the particles (except the one
at $(0,0)$) from the CF Landau
level state $(n,-n)$ to $(n-1, -n+2)$. There are thus $N-1$ different ways
of constructing the $L=2$ CF Slater determinant which are all degenerate
in CF kinetic energy. In mathematical terms, this corresponds to replacing
one of the states $\eta_{n,-n} = \partial^n$ by $z\partial^{n-1}$ (see \pref{etanl}).
It thus looks as if there are $N-1$ different CF candidate wave functions
for the $L=2$ state, while according to the above arguments, we would expect 
the CF wave function to be unique and exact! 
The solution to this apparent paradox is that {\em while the $N-1$ possible
Slater determinants are certainly not equal to each other, they all
result in the same polynomial after acting on the Jastrow factor.}
To see how this comes about, consider again the
case of four particles. The first possible Slater determinant is
\be
f_S=
 \left| \begin{array}{cccc}
z_1 & z_2 & z_3 & z_4 \\
1 & 1 &  1 & 1 \\
\partial_1^2 & \partial_2^2 & \partial_3^2  & \partial_4^2 \\
\partial_1^3 & \partial_2^3 & \partial_3^3  & \partial_4^3
 \end{array} \right|
\ee
resulting in
\be
f_U&=& \sum_{\sigma} \e(\sigma) z_1 z_2^0 \p_3^2 \p_4^3 \times
                         z^{\sigma(1)}_1 z^{\sigma(2)}_2 z^{\sigma(3)}_3 z^{\sigma(4)}_4 \\
	&=&\sum_{\sigma} \e(\sigma) z_1^{\sigma(1)+1} \cdot z_2^{\sigma(2)} 
		\cdot \sigma(3)(\sigma(3) - 1)  z_3^{\sigma(3)-2}  \nonumber \\
		&\cdot& \sigma(4)(\sigma(4) - 1)(\sigma(4)-2)  z_4^{\sigma(4)-3} 
\ee
which equals the determinant
\be
f_U=
 \left| \begin{array}{cccc}
z_1 & 1 & 0 & 0\\
z_1^2 & z_2 & 0 & 0\\
z_1^3 & z_2^2 & 2 & 0\\
z_1^4 & z_2^3 & 6z_3 & 6 
 \end{array} \right|
= 12 (z_1 z_2 - z_1^2).
\ee
Proceeding in the same way for the other two possibilities gives
$f_U = 12(z_3^2 - z_2 z_3)$ and $f_U = 12(z_4^2 - z_3 z_4)$, respectively.
Thus, since the {\em full} wave function is constructed from $f_U$ by
summing over all permutations of the indices (1,2,3,4), the final,
symmetrized sum will be the same for all these constructions.

Exactly the same thing happens for $L=3$. There are $N-2$ ways of
constructing the Slater determinant, obtained from the $L=0$ one by 
letting any $\p^n \rightarrow z \p^{n-2}$. Again, these different constructions
simply correspond to picking out different parts of the final, symmetrized
polynomial.

Of course, similar things may happen at other angular momenta as well.
For example, there are two ways of constructing the Slater determinant
for $L=N-1$. Again, these result in the same polynomial, though in
this case the CF construction does not give the exact wave function
even in the large $N$ limit.

The fact that the wave functions resulting from the CF construction
for $L=2,3$ are equal to the very simple and compact expressions in
\pref{TI1}, again provides non-trivial and potentially useful mathematical
identities.

\section{Summary and outlook}
To summarize, we have shown that the CF construction for bosons at
low angular momenta may be taken to far higher particle numbers
than previously by handling the derivatives of the CF Slater determinant
in a new way. In particular, we used this to show 
that for the single vortex, $L=N$, the CF construction becomes
exact in the high $N$ limit, leading to a non-trivial and potentially
useful mathematical identity. For other $L$, the CF 
construction does, in general, {\em not} produce the exact wave functions.
We have argued why it is expected to give the exact ground state
for $L=2,3$ (all $N$). However, work remains to get a better understanding
of what makes $L=N$ special.
Moreover, we used our methods to show examples
of other mathematical peculiarities of the CF construction in the
regime $L < N$. Our hope is that the results of this paper may
provide new insight into properties of CF wave functions in the 
few-vortex regime beyond $L=N$, where no exact, analytical 
wave functions are known. For example, a preliminary study of the case 
$L=2N$ suggests that while the number of possible CF Slater determinants 
increases linearly with $N$, these may reduce to just two different wave
functions; finding the ground state as a linear combination of these would
then just be a one-parameter problem. There is thus hope that one may construct 
rather simple analytical trial wave functions in the regime $L \sim N$,
which may directly reveal the vortex structures of these states.
The present paper indicates that in this
regime, the CF construction may do far better than previously
assumed in the literature.

\vskip 2mm
\noi {\bf Acknowledgement}: We thank T.H. Hansson and J.K. Jain for 
useful discussions. This work was supported in part by the Institute for Theoretical  
Sciences, a joint institute of Argonne National Laboratory and the University of
Notre Dame, funded through DOE contract W-31-109-ENG-38 and Notre Dame
Office of Research.  Financial support from NordForsk is gratefully
acknowledged.

\vspace{-0.5cm}
\bibliographystyle{unsrt}

\begin{thebibliography}{10}

\bibitem{baym1} For a nice review, see G. Baym, cond-mat/0408401 and references therein.

\bibitem{madison1} K.W. Madison, F. Chevy, V. Bretin, and J. Dalibard,
Phys. Rev. Lett. {\bf 86}, 4443 (2001).

\bibitem{abo1} J.R. Abo-Shaeer, C. Raman, J.M. Vogels, and W. Ketterle,
Science {\bf 292}, 476 (2001).

\bibitem{haljan1} P.C. Haljan, I. Coddington, P. Engels, and E. Cornell,
Phys. Rev. Lett. {\bf 87}, 210403 (2001).

\bibitem{engels1} P. Engels, I. Coddington, P.C. Haljan, and E.A. Cornell,
Phys. Rev. Lett. {\bf 89}, 100403 (2002).

\bibitem{schweickhard1} V. Schweickhard, I. Coddington, P. Engels, V.P. Morgendorff,
and E.A. Cornell,
Phys. Rev. Lett. {\bf 92}, 040404 (2004).

\bibitem{wilkin1} N.K. Wilkin, J.M.F. Gunn, and R.A. Smith,
Phys. Rev. Lett. {\bf 80}, 2265 (1998).

\bibitem{wilkin2} N.K. Wilkin and J.M.F. Gunn,
Phys. Rev. Lett. {\bf 84}, 6 (2000).

\bibitem{cooper1} N.R. Cooper and N.K. Wilkin, 
Phys Rev. B {\bf 60}, R16279 (1999).

\bibitem{jainreview} J.K. Jain and R.K. Kamilla, in 
{\em Composite Fermions: A unified view of the quantum Hall effect},
edited by O. Heinonen (World Scientific, River Edge, NJ, 1998).

\bibitem{viefers1} S. Viefers, T.H. Hansson, and S.M. Reimann,
Phys. Rev. A {\bf 62}, 053604 (2000). 

\bibitem{chang1} C. Chang, N. Regnault, T. Jolicoeur, and J.K. Jain,
Phys. Rev. A {\bf 72}, 013611 (2005).

\bibitem{regnault1} N. Regnault and Th. Jolicoeur,
Phys. Rev. Lett. {\bf 91}, 030402 (2003);
Phys. Rev. B {\bf 69}, 235309 (2004).

\bibitem{bertsch99} G.F. Bertsch and T. Papenbrock,
Phys. Rev. Lett. {\bf 83}, 5412 (1999).

\bibitem{smith1} R.A. Smith and N.K. Wilkin, 
Phys. Rev. A {\bf 62}, 061602 (2000).

\bibitem{jaincomm} J.K. Jain, private communication.

\bibitem{qdcf} J.K. Jain and T. Kawamura, Europhys. Lett. {\bf 29}, 321 (1995);
B. Rejaei, Phys. Rev. B {\bf 48}, 18016 (1993);
C.W.J. Beenakker and B. Rejaei, Physica B {\bf 189}, 147 (1993).

\bibitem{kjonsberg1} H. Kj\o nsberg and J.M. Leinaas,
Nucl. Phys. B {\bf 559}, 705 (1999).

\bibitem{jeon1} G.S. Jeon, K.L. Graham, and J.K. Jain,
Phys. Rev. Lett. {\bf 91}, 036801 (2003).

\bibitem{trugman1} S.A. Trugman and S. Kivelson,
Phys. Rev. B {\bf 31}, 5280 (1085).
                          

\bibitem{footnote} There is a class of wave functions within the CF formalism 
that are, by construction, translation invariant:
They are called {\em compact states} and are characterized 
by having the $n$th CF Landau level occupied from $l_n = -n$ to $l_n = l_n^{max}$
without any ``holes''.

\end{thebibliography}

\vspace{0.5cm}
\end{document}